# Randomizing genome-scale metabolic networks

## Areejit Samal[1,2] and Olivier C. Martin[3,1*]

[1] Laboratoire de Physique Théorique et Modèles Statistiques (LPTMS), CNRS and Univ Paris-Sud, UMR8626, Orsay Cedex, France
[2] Max Planck Institute for Mathematics in the Sciences, Leipzig, Germany
[3] INRA, UMR0320/UMR8120 Génétique Végétale, Univ Paris-Sud, Gif-sur-Yvette, France

*Correspondence should be addressed to: Olivier C. Martin (olivier.martin@u-psud.fr)

Email addresses: AS: samal@mis.mpg.de; OCM: olivier.martin@u-psud.fr

## Abstract

Networks coming from protein-protein interactions, transcriptional regulation, signaling, or metabolism may appear to have "unusual" properties. To quantify this, it is appropriate to randomize the network and test the hypothesis that the network is not statistically different from expected in a motivated ensemble. However, when dealing with metabolic networks, the randomization of the network using edge exchange generates fictitious reactions that are biochemically meaningless. Here we provide several natural ensembles of randomized metabolic networks. A first constraint is to use valid biochemical reactions. Further constraints correspond to imposing appropriate functional constraints. We explain how to perform these randomizations with the help of Markov Chain Monte Carlo (MCMC) and show that they allow one to approach the properties of biological metabolic networks. The implication of the present work is that the observed global structural properties of real metabolic networks are likely to be the consequence of simple biochemical and functional constraints.

## Introduction

Social networks exhibit "small world" characteristics [1,2]; food webs have hierarchies coming from trophic levels [3]; gene networks have small in-degree, broad out-degree, and contain strongly over-represented motifs [4]. These kinds of "remarkable" features distinguish natural or even human made networks from random graphs [5,6,7]. However, comparing the networks arising in these different systems to random graphs is unsatisfactory because it ignores all potentially relevant underlying factors that constrain these networks. One should also ask whether these networks are remarkable given where they come from, taking into account the known factors which shape them. A way to address this issue is to perform *graph randomization*. The most commonly used such approach for biological networks is based on performing edge exchanges [4,8,9]. This algorithm (illustrated in Figure S1) by construction preserves the network's degree distribution exactly.

     Our focus in the present work is on metabolic networks. Previous studies have revealed that metabolic networks of living organisms are highly structured. For example, the degree distribution of the metabolites in these networks has a power law tail [10,11]. Metabolic networks seem to have further remarkable features such as a high level of clustering [12]. However, to claim that such features are remarkable, one has to use a benchmark. The use of random graphs has the drawback of ignoring the special nature of the degree distribution. If instead the comparison is made using the edge exchange algorithm, one is confronted with a serious conceptual problem that is specific to metabolism: the randomized ensemble contains meaningless reactions (cf. Figure S1). That is because the edge exchange procedure ignores all biochemistry, and in particular the fact that most biochemical reactions correspond to adding or removing small groups. Furthermore, this naive randomization corresponds to using "random" reactions which will not balance mass, charge and even less atomic elements; clearly this is enough to cast a doubt on the relevance of such a procedure. To overcome this problem, we have little choice but to force the reactions to have a minimum of realism; that can be done by using reactions known to arise in various organisms or *in vitro*. This corresponds to the first level of "constraints" that should be imposed when randomizing metabolic networks.

Other levels can be introduced based on *functionality*. For instance, to understand the differences between the metabolic network of a given organism and "what might have been expected" in other realizations, one may appeal to the fact that organisms are alive, eat, reproduce *etc*. The purpose of the present work is to show how, within metabolic networks, one may introduce *randomized ensembles*; these ensembles can be used as benchmarks, allowing one to meaningfully ask whether a given organism's metabolic network is particularly atypical.

The outline of the paper is as follows. We first examine the fat tail in the metabolite degree distribution when using realistic biochemical reactions and investigate the source of this tail. Then we address the randomization problem in metabolic networks and introduce network ensembles subject to increasing levels of constraints. We study the structural properties of metabolic networks in these ensembles, including the clustering coefficient and sizes of the strong components. These results are discussed in the following section, while detailed methods are provided in the last section.

## Results

### Degree distribution and the KEGG_Hybrid set of reactions

Given a metabolic network such as illustrated in Fig. 1(a), the degree of a metabolite is the number of reactions in which it participates. With the availability of genome-scale metabolic networks [10], the metabolite degree distributions in a number of organisms have been determined. A striking result is that all organisms show metabolite degree distributions with fat tails well described by a power fall-off [10,11]. Furthermore, the power of this tail varies very little from one organism to another, being always close to 2.2 [10]. Clearly there exist metabolites involved in a large number of reactions; examples include ATP (which provides the transfer of a phosphate group), NADH (which provides the transfer of electrons) *etc*. This behavior is in fact typical of all metabolites of high degree: they transfer small groups and are therefore generally referred to as "currency" metabolites [13,14]. Because these currency metabolites arise in so many reactions, one can expect nearly all living organisms to produce them. If this occurs, one also expects a similar power law to arise in the metabolite degree distribution in different organisms.

To address this point in a quantitative framework, we ask what would be expected in a "random" organism, that is, in one using *random biochemical reactions*? One could introduce artificial reactions in which randomly chosen metabolites would be transformed into others; however this would not preserve atomic species, and even if one could enforce conservation, it would nearly always lead to reactions which have no existence. A more suitable approach is to restrict ourselves to biochemically realizable reactions. We used a database of such reactions compiled by Rodrigues and Wagner [15]. These authors combined the reactions in the Kyoto Encyclopedia of Genes and Genomes (KEGG) database [16] with those of the *E, coli* iJR904 metabolic model [17] and curated the resulting set (see Materials and Methods). This leads to a total of 5870 reactions and 4816 metabolites; we shall refer to this list of reactions and metabolites as KEGG_Hybrid. Given this list of possible reactions and metabolites, Fig. 2 shows the degree distribution of the metabolites within the database on a log-log scale. The power law is clearly a good approximation; fitting these data using the method in [18,19] gives an exponent of 2.31. This value is close to the exponent for the *E. coli* genome-scale metabolic network [17] which is 2.17; the corresponding distribution is also displayed in Fig. 2.

The metabolic network of *E. coli* has far fewer reactions than the 5870 in the KEGG_Hybrid database, and so the maximum degree in *E. coli* must be smaller; this is visible in Fig. 2. Furthermore, our objective is to compare *E. coli* to "random" organisms so we should force the number of reactions to be the same as in *E. coli*, allowing any of the biochemical reactions in KEGG_Hybrid. This defines a simple *ensemble* of possible metabolic networks where the biochemical constraint of using real reactions is enforced. We have thus generated 1,000 random genomes (lists of $n$ reactions chosen at random in KEGG_Hybrid, $n=n_E=831$ being the number of reactions of the *in silico E. coli* metabolic network) and computed the degree distribution in this ensemble. The result is displayed in Fig. 2 with the label "Random", and the distribution again seems to follow a power law but with a slightly higher exponent, around 2.51. Thus this "Random" ensemble leads to metabolic networks whose metabolite degree distribution has characteristics rather similar to those of *E. coli*.

The similarity of the three distributions in Fig. 2 may seem remarkable, but upon reflection it can be understood as follows. The highest degree metabolites in the KEGG_Hybrid set participate in many reactions. Thus they are most likely very important biochemically, so they should be present in *E. coli*. Quantitatively, we have checked this: among the metabolites of degree at least 50 in KEGG_Hybrid, 94% are also present in *E. coli*. Furthermore, this same pattern is expected in the Random ensemble simply because choosing reactions at random gives a higher probability of incorporating metabolites that participate in many reactions. Again this can be tested explicitly: for any of the metabolites that have degree at least 50 in KEGG_Hybrid, the Random metabolic networks include them with probability above 0.99. Interestingly, the biochemical nature of these high degree metabolites is quite specific: they are categorized as "carriers" or as "precursors" in the biochemical literature [13]. Tanaka and Doyle [14,20] have investigated the degree properties of these two classes and found that indeed they are the contributors to the fat tails in the degree distribution for different organisms, but we see here that this also holds for the KEGG_Hybrid set and for our "Random" ensemble.

All the above concerns the degree of the metabolites. The same kind of analysis can be performed for the degree of the *reactions*, where the degree of a reaction is given by the number of its substrates (metabolites it involves). In contrast to metabolites, reactions do not have high degrees: a typical reaction will involve just a few metabolites, the most frequent number being 4, and very rarely will there be more than 6. This situation is illustrated in Fig. 3 where we also distinguish the different kinds of metabolites (see Materials and Methods for details). One sees from Fig. 3 that reactions typically involve a currency pair, sometimes two, but that almost always there are at most 3 metabolites per reaction that are *not* of the currency type.

## Ensembles: from implementing biochemical realism to allowing for functional constraints

The "Random" ensemble is a first way to define randomized metabolic networks: it takes into account both the need to use meaningful biochemical reactions and the number of reactions in the genome of interest. We shall consider that the metabolic network (of an organism or of a randomized organism) is specified by its set of enzyme coding genes, and we shall refer to this set as its "genotype".

The approximate power law tail found in the degree distribution of metabolites in living organisms can be traced to the contribution of currency metabolites (cf. Fig. 2). However, numerous other statistical properties of biological metabolic networks set them aside from those in the Random ensemble; for instance their number of metabolites is significantly lower, they have fewer "blocked" [21,22] reactions (reactions that cannot sustain flux for instance because they are not connected to other reactions), *etc*. Just as we fixed the number of reactions *n* in the genotypes forming the Random ensemble, it is appropriate to include further "macroscopic" constraints to refine the randomization ensembles. For each added constraint, one can expect the statistical characteristics in the ensemble to become closer to those of living organisms, but the *hope* is that just a few relevant constraints will be sufficient to have a quite satisfactory randomized ensemble. Beyond the constraints already mentioned, there is the simple fact that metabolism of living organisms allows them to grow and reproduce. Although these are incredibly complex tasks, genome-scale metabolic network models [23,24] take into account the realizable fluxes through biochemical reactions and the possibility that a given set of reactions (catalyzed by enzyme coding genes) may produce all the biomass components necessary for cell growth. In our most constrained ensemble, we shall thus enforce the "functional" constraint that the genotype's metabolic network allows for production of these biomass components. The simplest ensemble is the one previously introduced under the term "Random" and we shall label it *R* because it is simply based on using a fixed number of random *reactions* in KEGG_Hybrid. Adding the constraint of the number of *metabolites* gives the ensemble we label *RM* and so forth. We now describe these successive ensembles and the computational tools used to sample them. We shall then examine the statistical properties of the networks in each of these ensembles and see the effects of successively adding these constraints.

*The ensembles R, RM, and uRM*

The allowed reactions will always be taken from the aforementioned KEGG_Hybrid list. This guarantees that every reaction satisfies atomic conservation laws. Also, these reactions are either reversible or irreversible, and this is taken into account in the modeling. The first ensemble *R* constrains the number of reactions; it consists of genotypes having exactly *n* reactions in the KEGG_Hybrid list where *n* is the number of reactions of the "reference" organism which one wants to benchmark in a randomization test; for specificity, the reader can think of this reference organism as being *E. coli*. The second ensemble further constrains the number of metabolites; the number of metabolites *m* in a genome is obtained by counting all the distinct metabolites associated with the reactions in that genotype. In practice, the sampling is simpler if one constrains *m* to be in a range; we shall use $m \leq m_E$ where $m_E$ is the number of metabolites in the reference organism *E. coli*. We denote this second ensemble which imposes two constraints by *RM*. (Note that to go from a sample of $m \leq m_E$ to one with $m = m_E$, it is sufficient to use the subsample satisfying $m = m_E$.) The third ensemble, denoted by *uRM*, restricts the nature of the reactions in KEGG_Hybrid that we permit ourselves to consider. The motivation for this restriction comes from the fact that KEGG_Hybrid includes many reactions involving "exotic" metabolites which are involved with just one reaction. In such cases, those reactions will necessarily be isolated and thus "afunctional" biochemically; unless the associated metabolites are part of the biomass, such reactions would have no reason to be kept in a biological organism. (Note that some of these cases may be due to errors or missing reactions in KEGG_Hybrid; these limitations are expected to be resolved in the not so distant future as databases get improved.) A similar situation arises when a reaction requires a metabolite that can only be produced in particular chemical environment that we do not consider; such a reaction will then be blocked. Our working definition of a "blocked" reaction is based on the possible fluxes it can sustain in the steady state; if a reaction is guaranteed to be never used in such conditions, then it is considered as blocked and removed from the KEGG_Hybrid list. In practice, for all the reactions in KEGG_Hybrid we determine whether they are blocked [21,22] (cannot sustain non zero flux; see Materials and Methods for the details); the reduced set of *unblocked* reactions is then used as input for constructing genotypes. The nomenclature *uRM* of the ensemble indicates that we use unblocked reactions with constrained numbers of reactions *and* metabolites.

### *The ensembles uRM-V1, ..., uRM-V10*

The essence of living organisms is growth and reproduction. Metabolism plays a central role therein, transforming various nutrients brought in from outside the cell into primary metabolites (amino acids, nucleic acids, fatty acids, *etc*). These are then used as bricks for building proteins, DNA, lipids, *etc*. Genome-scale metabolic models provide a tested framework to relate genotypes (lists of enzyme-coding genes) to metabolic capabilities and phenotypes [23,24]. The framework, often called FBA for "Flux Balance Analysis", allows one to compute the possible flux distributions through all the reactions assuming the metabolic network is in the steady state [23,24,25]. One may ask whether all the biomass compounds can be produced given a chemical environment, *i.e.,* a set of nutrients defining allowed input fluxes into the cell. If a genotype's metabolic network satisfies this constraint, we say that the genotype is "viable" on that chemical environment because the *in silico* FBA predicts that the cell can grow given those nutrients. This is illustrated in Fig. S2.

There are many possible choices of nutrients; clearly one needs sources of all major elements (H, C, O, N, S, P). It is common practice to focus on the carbon utilization because it is often limiting; we thus work with "minimal" environments having a single carbon source. In the lab, it is easy to test whether a microorganism grows on a whole panel of different environments, and the corresponding list of growth/no-growth results is referred to as the *growth phenotype* of the organism. This growth phenotype can be considered a constraint to impose on a randomization. In this spirit, we consider a succession of ensembles associated with viability on multiple environments. Specifically, given an organism like *E. coli* and its growth phenotype, we can consider the random genotypes that have the same *in silico* growth phenotype (as predicted by FBA). The associated ensemble thus takes into account viability constraints. These constraints can be considered as being imposed successively: one can force growth first on one chemical environment, then on two, then on three, and so on. We refer to these ensembles as *uRM-V1*, *uRM-V2*, *uRM-V3*, etc. Interestingly, the first step, namely going from *uRM* to *uRM-*

*V1,* turns out to be the most stringent as the ones thereafter give rise to only rather small changes.

*MCMC sampling of each ensemble*

The ensemble $R$ can be sampled by drawing genotypes with the correct number of reactions, but the constraints inherent to the other ensembles do not allow such a simple procedure. We thus resort to Markov Chain Monte Carlo (MCMC) as a way to sample each ensemble. This requires obtaining an element of the ensemble as a starting point, and then performing random walks within the ensemble. Each trial step involves doing a reaction swap (exchanging a reaction in the genotype with one that is not) to respect the constraint of having a fixed number of reactions. Then the different constraints of the ensemble of interest are checked; if they are satisfied, the trial step (to a new genotype) is accepted, otherwise it is rejected. The schematic representation of this is given in Fig. 4 while the detailed procedures are given in Fig. S3 for the ensemble *uRM-V1*. Note that by construction, in each ensemble, all of the elements are equiprobable; our ensembles are then just nested sets of genotypes satisfying increasing numbers of constraints. The reduction in size of these sets as each constraint is added can be extremely severe. For instance, when going from all reactions to unblocked reactions, over half of the reactions are discarded, leading to a reduction of the genotype space by approximately a factor $2^n$. Similarly, it was shown in previous work [26] that including the viability constraint on the first chemical environment leads to a reduction by at least a factor $10^{22}$. These numbers drive home the necessity of using MCMC for sampling: direct random sampling is hopelessly inefficient.

## Metabolic network statistical properties

*Genetic diversity*

In any of our ensembles sampled by MCMC, two genotypes taken at random will share some reactions and differ in others. For instance, in the ensemble *uRM-V10,* we find a total of 106 specific reactions that are *necessary* for any genotype to have nonzero biomass flux. These 106 reactions are then present in every genotype of the ensemble *uRM-V10*. Nevertheless, two genotypes taken at random in this ensemble tend to be rather different. Specifically, if one takes two genotypes $G_1$ and $G_2$ at random in *uRM-V10*, we find that on average $G_2$ will have more than 50% of its reactions that are not in $G_1$. This also holds true when we compare a random genotype to that of *E. coli*. In the ensembles with fewer constraints (eg. *uRM-V1*), the level of dissimilarity between random genotypes is even higher. Thus, in spite of some shared reactions in all genotypes, the genotypes in our ensembles are not very similar to each other or to *E. coli*: the ensembles have a high level of genetic diversity.

*Global topological properties*

To each genotype is associated a list of reactions and their corresponding metabolites; the whole can be represented by a bipartite graph (cf. Fig. 1(a)). From this graph, one may form a reduced graph for only the metabolites, or one for only the reactions; these graphs are called the metabolite-metabolite graph and the reaction-reaction graph respectively. (See the Materials and Methods section for the associated procedures.) It is appropriate to emphasize that the bipartite graph representation of the metabolic network contains more information than its associated unipartite graph representation.

For each genotype generated and saved in the different ensembles, we have constructed its metabolite-metabolite graph. Then we measure several of the standard *structural* properties of that (directed) graph. These are as follows. (1) The clustering coefficient $C$ which roughly is a measure of the frequency of triangles in the network. (2) The average path length $L$ between randomly chosen nodes. (3) The probability $P_C$ that two randomly selected nodes A and B are connected by a directed path from A to B; this gives an indication of whether a metabolite is involved in another's production. (4) The size of the largest strong component (LSC) which measures the connectivity of the network. In a directed graph, a strong component is defined as a maximal sub-graph such that there exists a directed path between any two of its nodes; in the case of an undirected graph, it is then just a maximal connected component. We focus on the largest of these strong components in this work. (5) The "IN" (respectively the "OUT") sub-graph for a given strong component is the set of nodes for which there is a directed path to (respectively from) the strong component [27]. We shall monitor the union of the largest strong component (LSC) and its associated IN and OUT parts.

For each of these indicators of graph structure, we have computed the mean values within the different ensembles for the metabolite-metabolite graph, and have also determined the value for the graph associated with the *E. coli* genotype. In Fig. 5 we display as a function of the increasingly constrained ensembles three structural quantities related to connectivity: the average of $P_C$, the average size of the LSC, and the average size of LSC+IN+OUT. To the right of the bar associated with *uRM-V10* we show the value for *E. coli*. Clearly the first constraints strongly affect the structure of the metabolic networks, while increasing the number of environments on which one forces viability gives rise only to modest changes. When considering the other structural properties of networks in the ensembles, we see that for the clustering coefficient, the first constraint (going from *R* to *RM*) is the most important (cf. Fig. S4). For the average path length, already at the level of *R* one has quite good agreement with the value in *E. coli*, just as was the case for the degree distribution (cf. Fig. S5). Thus we have ensembles of randomized metabolic networks that are good benchmarks of comparison for the biological network.

*Functional constraints shape global network structure*

The trends described above can be summarized by following the joint statistics of the structural properties in the ensembles as one adds successive constraints. This is illustrated in Fig. 6 for three of the structural properties. Each ensemble is represented by 1000 of its genotypes and for clarity we have displayed only three of the ensembles. We see a systematic change in the structural properties as constraints are added, and that the three clouds associated with the constraints represented here have little overlap. Note that by construction our ensembles are actually embedded sets of genotypes; each added constraint reduces the genotypes allowed. Such a hierarchical structure does not prevent the clouds in Fig. 6 from being rather well separated. Furthermore, the trend with addition of constraints very clearly brings the clouds closer to the point representing the structural properties of *E. coli*. This was visible too at the level of the individual observables (cf. Figs. 5 and 6).

While computing the above mentioned structural properties, the construction of the metabolite-metabolite graph plays a key role. However, the unipartite metabolite graph construction relies on a classification of metabolites into currency and non currency metabolites, and such a classification is not clear-cut. Indeed some currency metabolites have a carrier role in some reactions and a non carrier role in others. To check that our conclusions are not sensitive to some level of arbitrariness in the classification scheme, we have repeated the whole calculation for a modified set of currency metabolites, removing 20 currency metabolites from the original list in Table S1. We display in Fig. S6 the analog of Fig. 6; the difference between the two figures is hardly detectable by eye, showing that the trends found above are robust.

## Discussion

Over the past decade, genome-scale metabolic networks have been constructed for several organisms. In all cases, the metabolite degree distribution exhibits a fat tail compatible with a power decay [10,11] of exponent around 2.2. We found that this behavior can be traced to the metabolite degree distribution in the set of all known biochemical reactions as given for instance in KEGG. The fundamental source of these fat tails is the large number of *currency* metabolites that transfer small groups in nearly all real biochemical reactions. It is essential to take into account this fact when testing whether biological metabolic networks have unexpected features. The commonly used edge exchange algorithm has the desirable property of preserving the network degree distribution but it is inappropriate because the procedure introduces fictitious reactions having no meaning. Any sensible testing framework should force the benchmark (the randomized ensemble) to incorporate real biochemical reactions. We showed that this could be done in practice by using a database of real biochemical reactions compiled from KEGG and iJR904. In this framework, we found that the degree distribution of metabolites had a fat tail very similar to what is seen in real organisms.

One may ask whether the observed fat tail is an artifact of the KEGG database itself which summarizes the reactions in today's organisms. It is quite possible that other reactions and cofactors can act as substitutes to the ones occurring in KEGG, and thereby affect the degree distribution. However, it is likely that selection pressures act against such substitutes, for instance because of efficiency of catalysis or availability of molecular species. In effect, the use of KEGG reflects evolutionary constraints in addition to the purely biochemical ones. Thus, the present work is relevant for natural organisms but much less so for synthetic ones.

Another caveat associated with this study is the bias arising due to incompleteness of the KEGG database. Firstly, KEGG is missing transporters of less studied organisms; as a consequence a number of reactions appear to be blocked. Secondly, many biosynthesis pathways are incomplete; this is especially true for rare (or poorly understood) pathways. However, both of these biases can be expected to have only mild consequences within our study. Indeed for our choices of chemical environments, the curation of genome-scale models of different organisms has filled the gaps for transporters. Furthermore, our use of the *E. coli* biomass reaction formula makes our growth phenotypes insensitive to missing reactions as long as they arise in rare biosynthetic pathways.

Looking at structural properties beyond the degree distribution, we found that the ensemble *R* showed significant differences with the biological case (where the ensemble *R* corresponds to choosing randomly a given number of reactions in the database KEGG_Hybrid). Since understanding the topological properties of networks can give insights into their structure-function relationship, it is appropriate to refine the benchmark ensemble. Thus, we successively added further global constraints, in particular by enforcing metabolic capabilities, in this context biomass production. Adding such *functional* constraints takes into account the growth properties of living organisms and thus the "macroscopic" forces which shape biological metabolic networks. We find that by adding biochemical and functional constraints, the structural properties of the random networks in our ensembles become very close to what is seen biologically as illustrated in Fig. 6; that this is possible without taking into account any microscopic properties is really remarkable. Depending on the structural feature considered, we find that some features emerge relatively "early", that is follow from fewer macroscopic constraints than others.

Perhaps most strikingly, these trends occur within ensembles that maintain a high level of genetic diversity. Indeed even in our most constrained ensemble, *uRM-V10*, the metabolic networks show large differences in reaction usage. Quantitatively, two randomly chosen networks in the ensemble *uRM-V10* will typically differ in half of their reaction content. As a cautionary note, it is important to stress that the observed trends here concern global structural measures commonly used in general network analysis. One cannot exclude the possibility that consideration of metabolism-specific observables based for instance on fluxes may lead to a different picture.

In conclusion, the present work indicates that the observed global structural properties of metabolic networks in living organisms are likely to be consequences of the simplest biochemical and functional constraints. Such a possibility has been previously suggested [28,29] but remained in the spirit of a conjecture; we hope that the direct computational evidence provided in this work will transform conjecture into paradigm.

## Materials and Methods

### Biochemical reaction sets

#### KEGG_Hybrid reaction set

We have used a hybrid database compiled by Rodrigues and Wagner [15] containing 4816 metabolites and 5870 biochemical reactions for this work. This database of 5870 reactions was compiled by merging the Kyoto Encyclopedia of Genes and Genomes (KEGG) LIGAND reaction database [16] with the *E. coli* genome-scale metabolic model iJR904 [17], followed by appropriate pruning to exclude elementally imbalanced and generalized polymerization reactions [15]. Of the 5870 reactions in the hybrid database, 3369 are irreversible and 2501 are reversible reactions. Also, more than 5500 reactions are contained in the KEGG LIGAND database and so less than 300 reactions are specific to the *E. coli* genome-scale metabolic model iJR904. In this work, we will refer to the set of 5870 reactions contained in the hybrid database [15] as "KEGG_Hybrid reactions".

The hybrid database also contains transport reactions for 143 external metabolites in the *E. coli* iJR904 metabolic model; these can be used to transport such metabolites across the cell boundary. The 143 external metabolites were taken to be the set of possible uptake and secreted metabolites in the network. Further, an objective function Z in the form of a biomass reaction, that requires synthesis of all biomass components of *E. coli*, as defined in the iJR904 model [17], is also included in the hybrid database. The biomass reaction is used to determine the viability of a network.

*Unblocked KEGG_Hybrid reaction set*

Genome-scale metabolic networks typically contain "blocked" reactions that can have only zero flux in every investigated chemical environment under any steady state condition [21,22]. Such blocked reactions cannot contribute to the steady state flux distribution. With the set of 143 external metabolites in the *E. coli* iJR904 model, we found 2968 of the 5870 reactions in the hybrid database to be blocked under all environmental conditions [26]. We have excluded the 2968 blocked reactions from the set of 5870 reactions in the hybrid database to arrive at a reduced reaction set of 1597 metabolites and 2902 reactions. We refer to this reduced set of 2902 reactions in this work as "Unblocked KEGG_Hybrid reactions".

*The E. coli metabolic network*

The *E. coli* metabolic model iJR904 [17] contains 931 reactions which are also part of the hybrid database. After having excluded the 2968 blocked reactions from the hybrid database, the unblocked reaction set of 2902 reactions still contains 831 reactions of the *E. coli* iJR904 model. In this work, we refer to this set of 831 reactions as the *E. coli* metabolic network.

## Graph-theoretic representations of metabolic networks

The metabolic network can be represented as a directed bipartite graph built up of two types of nodes, metabolites and reactions, connected by two types of links. We can distinguish reactant metabolites from product metabolites of a reaction as follows: A link from a metabolite node to a reaction node specifies a reactant while a link from a reaction node to a metabolite node specifies a product. Note that in a bipartite graph, links between similar types of nodes are forbidden. It is important to differentiate between reversible and irreversible reactions in the network. In Fig. 1(a) we have used the bipartite representation to show three reactions in the glycolytic pathway.

Starting from a directed bipartite graph of metabolites and reactions, we can construct an associated directed unipartite graph of metabolites, referred to as a metabolite-metabolite graph. It summarizes the metabolic network structure by assigning links from reactant metabolites to product metabolites in each reaction. In the simplest definition, two metabolites will be "neighbors" (connected by a link) if and only if they appear in at least one common reaction [11]. However, a sizeable fraction of metabolites in the network have quite high degree, so this construction leads to very dense graphs whose statistical properties are dominated by the role of the currency metabolites. To overcome this problem and also maintain biochemical relevance, we construct the metabolite-metabolite graph by first removing the currency metabolites, and then assigning links from reactant metabolites to product metabolites in each reaction [13,14].

This representation has the advantage that the (directed) link between two metabolites signifies transformation of one into the other. For reversible reactions, the links between metabolites appear in both directions. See Fig. 1(b) for an illustration.

The different treatment of currency vs. non currency metabolites is based on the fact that biochemical reactions most often consist of adding or removing a small group (proton, phosphate, methyl, *etc*) of a large compound. Currency metabolites are the co-factors responsible for such transfers, and they are quite ubiquitous. Examples of currency metabolites include ATP, ADP, NADH, $NAD^+$, $H_2O$, $H^+$, Pi that are normally used as carriers for transferring electrons or certain functional groups such as phosphate group, amino group, methyl group, one carbon unit, etc. In our construction of the unipartite graph, we omit links arising due to presence of currency metabolites in each reaction. In Fig. 1(b), we show the unipartite graph corresponding to the bipartite graph shown in Fig. 1(a) for the three reactions in the glycolytic pathway. The list of currency metabolites used in our work was based on that in the paper by Ma and Zeng [13] and is given in Table S1.

## Structural properties of metabolic networks

*Metabolite degree distribution*

The degree of a metabolite $i$ (denoted by $k_i$) is the number of reactions in which the metabolite $i$ participates either as a reactant or a product in the network. The metabolite degree distribution $P(k)$ is defined as the probability that a randomly selected metabolite node participates in exactly $k$ reactions in the network. We use the bipartite graph representation of the metabolic network to compute the metabolite degree and degree distribution.

In Fig. 2, we have displayed several metabolite degree distributions after applying logarithmic binning. It is seen that the metabolite degree distributions approximately follow a power law, $P(k) \sim k^{-\gamma}$ [5], and the degree exponents $\gamma$ were extracted by using the maximum likelihood estimate method [18,19] recently proposed by Newman and colleagues rather than by fitting the binned data.

*Reaction degree distribution*

The degree of a reaction is the number of substrates that participate either as a reactant or a product in it. The reaction degree distribution $P(k)$ gives the probability that a randomly selected reaction has exactly $k$ substrates in it. We use the bipartite graph representation of the metabolic network to compute the reaction degree and degree distribution. Fig. 3 shows this distribution in the KEGG_Hybrid database.

*Clustering coefficient*

The clustering coefficient quantifies the extent to which the neighbors of a node in a graph are connected to each other [1]. The global clustering coefficient of a graph measures the fraction of triangles among the connected triples [30]. It is given by: $C = \dfrac{N_\Delta}{N_3}$ where $N_\Delta$ is the number of triangles and $N_3$ is the number of connected triples in the graph. In this work, we have computed the clustering coefficient for each network in our ensemble using the unipartite metabolite graph representation. Note that when computing the clustering coefficient, the graph is considered undirected.

*Path length and connectivity*

The average path length $<L>$ is a measure of the overall navigability in a network. It is defined as the average length of the shortest paths between all pairs of nodes in the directed unipartite metabolite graph. When computing the average path length for a disconnected graph, one considers only the node pairs for which a directed path exists. We have also computed the probability $P_C$ that a directed path exists between any two nodes in the unipartite metabolite graph. The clustering coefficient $C$, average path length $<L>$ and probability $P_C$ that a path exists between two nodes in a graph were computed using the igraph library [31].

*Largest strong component*

Given a directed graph, a strongly connected component is a maximal set of nodes such that for any pair of nodes $i$ and $j$ in the set there is a directed path from $i$ to $j$ and from $j$ to $i$ [32]. In general, a directed graph may have one or many strong components. The strong components of a graph are disjoint sets. The strong component with the largest number of nodes is designated as the *largest strong component* (LSC). The associated IN component consists of nodes which have access to LSC nodes via some directed path, but lack access from LSC nodes back to them via any directed path. The OUT component consists of nodes which can be accessed from the LSC nodes via some directed path, but lack access to LSC nodes from them via any directed path. Note that the so-called "bow-tie" architecture of networks is based on these LSC, IN and OUT components; that architecture has been observed both in the World Wide Web (WWW) [27] and in bacterial metabolism [33,34]. In this work, we have computed the fraction of nodes in the largest strong component (LSC) and in the union of LSC, IN and OUT components for networks in our ensembles using the directed unipartite metabolite graph representation.

## Genotype-to-phenotype map

A metabolic network genotype is any subset of reactions taken from the global reaction set itself consisting of $N$ reactions. A simple representation of a metabolic network genotype $G$ uses a binary string of length $N$, e.g., $G = (b_1,...,b_N)$, with each reaction $i$ being either present ($b_i$=1) or absent ($b_i$=0) (see Fig. S2 for an example). Each randomized network in our ensemble can be thought of as one genotype existing in a vast genotype space of possible metabolic networks. For any genotype, we can use flux balance analysis (FBA) [23,24,25] to determine whether the corresponding network has the ability to synthesize all biomass components in a given chemical environment or medium. FBA primarily uses information about the stoichiometry of reactions in the network to obtain a prediction for the steady-state fluxes of all reactions and the maximum possible biomass synthesis rate. The predictions of FBA and related approaches are generally in good agreement with experimental results [35,36].

We consider a genotype to be "viable" in a given chemical environment if and only if its maximum biomass flux predicted by FBA is non-zero (see Fig. S2). Otherwise, we consider the genotype to be non-viable. We use FBA and the *E. coli* biomass composition [17] to determine viability of a genotype in different chemical environments corresponding to minimal media. Specifically, we consider only minimal environments that contain a limited amount of a carbon source, along with unlimited amounts of the following inorganic metabolites: oxygen, water, protons, sulfate, ammonia, pyrophosphate, iron, potassium and sodium. Here, we have considered 10 carbon sources: glucose, acetate, succinate, pyruvate, oxoglutarate, glucose-6-phosphate, sucrose, acetaldehyde, glycerol and glycerol-3-phosphate.

## Generation of randomized ensembles

*Random ensemble R of genotypes with fixed number of reactions*

A genotype in the "random" ensemble $R$ can be simply generated by uniformly sampling subsets with exactly $n_E$ valid biochemical reactions from the KEGG_Hybrid reaction set of $N$=5870 reactions, where $n_E$ =831 is the number of reactions in the *E. coli* metabolic network. Using this procedure, we have generated 1000 genotypes in the random ensemble to compare with the *E. coli* metabolic network. Our motivation to fix the number of reactions in our genotypes is as follows: The biochemical reactions inside the cell are catalyzed by enzymes which are proteins coded by genes. By fixing the number of reactions in our genotype, we impose the constraint of fixed metabolic genome size.

*Ensemble RM of genotypes with fixed number of reactions and metabolites in the KEGG_Hybrid set*

The *E. coli* metabolic network consists of $n_E$ =831 reactions involving $m_E$ =668 metabolites. Though the genotypes in the random ensemble $R$ have exactly the same number of reactions as in *E. coli*, they typically contain many more metabolites than in the *E. coli* network. As a next step, we enforce the additional constraint that the genotypes have the same number of metabolites $m_E$ as in the *E. coli* network. Note that we cannot pick a fixed number of reactions at random from the KEGG_Hybrid reaction set if they are to satisfy the additional constraint of fixed number of metabolites. Hence, we use the Markov Chain Monte Carlo (MCMC) method to sample genotypes in the KEGG_Hybrid reaction set with same number of metabolites and reactions as in *E. coli*.

The MCMC method produces a sequence of genotypes forming a chain, the term "chain" coming from the property that the $(k+1)^{th}$ element of the sequence is generated from the $k^{th}$ one using a probabilistic transition rule. We start with the *E. coli* genotype and then propose a small modification in the genotype; if this modified genotype has its number of metabolites $\leq m_E$ (the number in *E. coli*), one accepts it as the next genotype of the sequence, otherwise the next genotype is identical to the current genotype. In this work, the modification introduced at each transition step is a reaction *swap*. That is, each modification adds one reaction from KEGG_Hybrid reaction set and removes another reaction from the current genotype, so as to keep the number of reactions $n_E$ constant in the genotype. The MCMC thereby produces a walk in the subspace of genotypes of $n_E$ reactions and at most $m_E$ metabolites. Starting from the initial *E. coli* genotype, we first carried out $10^5$ attempted swaps or Markov chain steps to erase the memory of the starting genotype. After this initial phase, we continued the MCMC procedure to sample genotypes with exactly $n_E$ reactions and at most $m_E$ metabolites. During this phase, it is not useful to keep all of the genotypes produced because they are strongly correlated. We thus saved only every $1000^{th}$ genotype generated, and we did $10^6$ steps. We refer to the set of 1000 genotypes with $n_E$ reactions and $\leq m_E$ metabolites within KEGG_Hybrid reaction set as the *RM* ensemble. We find that the procedure is relatively efficient, with an acceptance rate 0.22 that is not small. We also find that a substantial fraction of the networks have in fact $m=m_E$.

### *Ensemble uRM of genotypes with fixed number of reactions and metabolites in the Unblocked KEGG_Hybrid set*

"Blocked" reactions can have only zero flux in every investigated chemical environment under steady state conditions, and thus are "afunctional" in all genotypes. As a next step, we enforce the constraint that the genotypes in the ensemble are sampled within the Unblocked KEGG_Hybrid reaction set rather than the KEGG_Hybrid reaction set. We refer to this ensemble of genotypes containing the same number of metabolites $m_E$ and reactions $n_E$ as in the *E. coli* network within the unblocked reaction set as *uRM*. We generate the genotypes in the *uRM* ensemble through a slightly modified MCMC method from that mentioned above to generate the *RM* ensemble. In this case, at each transition step, we impose a reaction swap to the current genotype that is restricted to the Unblocked KEGG_Hybrid set, *i.e.,* we remove one reaction from the current genotype and add a reaction from the Unblocked KEGG_Hybrid set. The rest of the procedure is exactly the same as above for sampling *RM*. We have sampled 1000 genotypes in the *uRM* ensemble with $n_E$ reactions and at most $m_E$ metabolites.

### *Ensembles of viable genotypes with fixed number of reactions and metabolites*

The *E. coli* metabolic network has the ability to produce biomass components starting from nutrient metabolites in its environment for growth and maintenance. Thus as a next step, we enforce the additional functional constraint of growth in a chemical environment. We define the ensemble *uRM-V1* as that part of *uRM* in which the genotypes satisfy the functional constraint of non-zero biomass flux in the glucose minimal environment (as determined by FBA; see Fig. S2). We sample the genotypes in *uRM-V1* ensemble using a slightly modified MCMC method from that mentioned above to generate the *uRM* ensemble. In this case, at each transition step, we perform a reaction swap to the current genotype that is restricted to the Unblocked reaction set and accept the swap if the modified genotype satisfies the following two conditions: (a) the number of metabolites in the modified genotype is at most $m_E$, the number in *E. coli*, and (b) the modified genotype is viable under glucose minimal environment. The rest of the procedure is exactly same as when sampling genotypes in the ensemble *uRM*; a flowchart of the MCMC algorithm for sampling the ensemble *uRM-V1* is shown in Fig. S3. We have sampled 1000 genotypes in this *uRM-V1* ensemble.

Since, *E. coli* is able to survive and grow under diverse environmental conditions (rather than just one chemical environment), we have further generated two ensembles of genotypes satisfying increased functional constraints of (a) viability under 5 specified minimal environments (referred to as the ensemble *uRM-V5*) and (b) viability under 10 specified minimal environments (referred to as ensemble *uRM-V10*), respectively. These ensembles can be sampled by a MCMC procedure that is just a slight modification from the one shown in Fig. S3.

## Acknowledgements

We thank Alain Barrat, Pierre-Yves Bourguignon, Dominique de Vienne, Christine Dillmann, Silvio Franz, Sanjay Jain, Jürgen Jost, Marc Mézard and Andreas Wagner for constructive comments. We also thank João F. Matias Rodrigues and Andreas Wagner for sharing their database of reactions pooling those in KEGG and in iJR904. We also thank the anonymous reviewers for their constructive comments on the manuscript.

# References

1. Watts DJ, Strogatz SH (1998) Collective dynamics of 'small-world' networks. Nature 393: 440-442.
2. Milgram S (1967) The Small World Problem. Psychology Today 1.
3. Pauly D, Christensen VV, Dalsgaard J, Froese R, Torres F, Jr. (1998) Fishing down marine food webs. Science 279: 860-863.
4. Milo R, Shen-Orr S, Itzkovitz S, Kashtan N, Chklovskii D, et al. (2002) Network motifs: simple building blocks of complex networks. Science 298: 824-827.
5. Albert R, Barabasi AL (2002) Statistical mechanics of complex networks. Rev Mod Phys 74: 47-97.
6. Barabasi AL, Oltvai ZN (2004) Network biology: understanding the cell's functional organization. Nat Rev Genet 5: 101-113.
7. Newman M (2003) The structure and function of complex networks. Siam Review 45: 167-256.
8. Maslov S, Sneppen K (2002) Specificity and stability in topology of protein networks. Science 296: 910-913.
9. Guimera R, Sales-Pardo M, Amaral LA (2007) Classes of complex networks defined by role-to-role connectivity profiles. Nat Phys 3: 63-69.
10. Jeong H, Tombor B, Albert R, Oltvai ZN, Barabasi AL (2000) The large-scale organization of metabolic networks. Nature 407: 651-654.
11. Wagner A, Fell DA (2001) The small world inside large metabolic networks. Proc Biol Sci 268: 1803-1810.
12. Ravasz E, Somera AL, Mongru DA, Oltvai ZN, Barabasi AL (2002) Hierarchical organization of modularity in metabolic networks. Science 297: 1551-1555.
13. Ma H, Zeng AP (2003) Reconstruction of metabolic networks from genome data and analysis of their global structure for various organisms. Bioinformatics 19: 270-277.
14. Tanaka R (2005) Scale-rich metabolic networks. Phys Rev Lett 94: 168101.
15. Rodrigues JFM, Wagner A (2009) Evolutionary plasticity and innovations in complex metabolic reaction networks. PLoS Comput Biol in Press.
16. Kanehisa M, Goto S (2000) KEGG: Kyoto Encyclopedia of Genes and Genomes. Nucleic Acids Research: 27-30.
17. Reed JL, Vo TD, Schilling CH, Palsson BO (2003) An expanded genome-scale model of Escherichia coli K-12 (iJR904 GSM/GPR). Genome Biology 4: R54.
18. Clauset A, Shalizi CR, Newman MEJ (2009) Power-Law Distributions in Empirical Data. Siam Review 51: 661-703.
19. Newman M (2005) Power laws, Pareto distributions and Zipf's law. Contemporary Physics 46: 323-351.
20. Tanaka R, Csete M, Doyle J (2005) Highly optimised global organisation of metabolic networks. Syst Biol (Stevenage) 152: 179-184.
21. Burgard AP, Nikolaev EV, Schilling CH, Maranas CD (2004) Flux coupling analysis of genome-scale metabolic network reconstructions. Genome Res 14: 301-312.

22. Schuster S, Schuster R (1991) Detecting strictly detailed balanced subnetworks in open chemical reaction networks. J Math Chem 6: 17-40.
23. Feist AM, Palsson BO (2008) The growing scope of applications of genome-scale metabolic reconstructions using Escherichia coli. Nat Biotechnol 26: 659-667.
24. Price ND, Reed JL, Palsson BO (2004) Genome-scale models of microbial cells: evaluating the consequences of constraints. Nature Reviews Microbiology 2: 886-897.
25. Kauffman KJ, Prakash P, Edwards JS (2003) Advances in flux balance analysis. Curr Opin Biotechnol 14: 491-496.
26. Samal A, Rodrigues JFM, Jost J, Martin OC, Wagner A (2010) Genotype networks in metabolic reaction spaces. BMC Syst Biol: 30.
27. Broder A, Kumar R, Maghoul F, Raghavan P, Rajagopalan S, et al. (2000) Graph structure in the web. Computer Networks 33: 309-320.
28. Wagner A (2007) Gene networks and natural selection. In: Pagel M, Pomiankowski A, editors. Evolutionary Genomics and Proteomics. Sunderland, MA: Sinauer Associates Inc.
29. Papp B, Teusink B, Notebaart RA (2009) A critical view of metabolic network adaptations. HFSP J 3: 24-35.
30. Wasserman S, Faust K (1994) Social network analysis : methods and applications. Cambridge: Cambridge University Press. 825 p.
31. Csardi G, Nepusz T (2006) The igraph software package for complex network research. InterJournal Complex Systems: 1695.
32. Harary F (1969) Graph Theory: Addison-Wesley Publishing Company.
33. Csete M, Doyle J (2004) Bow ties, metabolism and disease. Trends Biotechnol 22: 446-450.
34. Ma HW, Zeng AP (2003) The connectivity structure, giant strong component and centrality of metabolic networks. Bioinformatics 19: 1423-1430.
35. Edwards JS, Ibarra RU, Palsson BO (2001) In silico predictions of Escherichia coli metabolic capabilities are consistent with experimental data. Nat Biotechnol 19: 125-130.
36. Segrè D, Vitkup D, Church GM (2002) Analysis of optimality in natural and perturbed metabolic networks. Proceedings of the National Academy of Sciences of the United States of America 99: 15112-15117.

# Figures

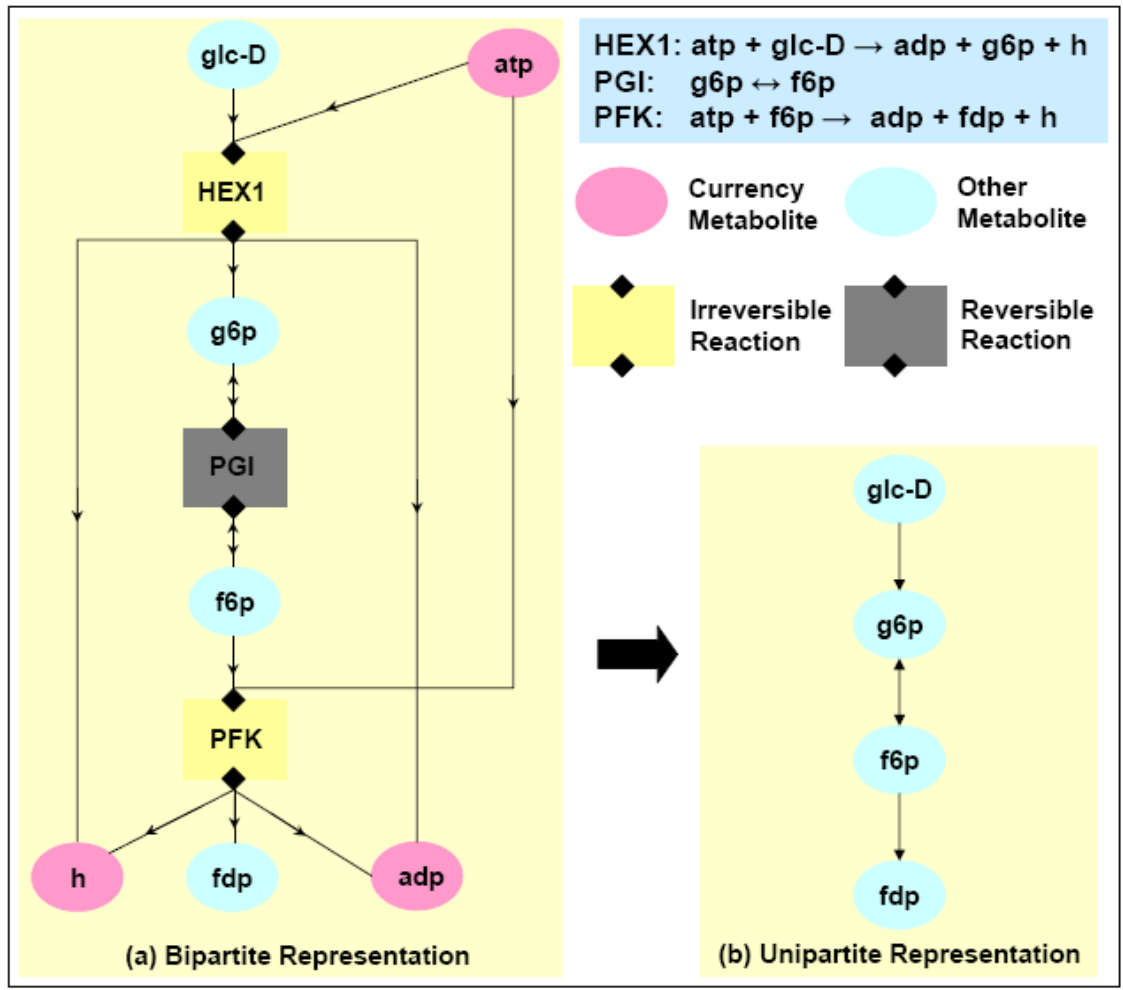

**Figure 1: Different graph-theoretic representations of a metabolic network.** (a) Bipartite graph representation for the three reactions, HEX1, PGI and PFK, in the glycolytic pathway. In the figure, reactions are depicted as rectangles and metabolites as ovals. Reversible reactions are shown in grey and irreversible reactions in yellow. The primary or other metabolites (cyan ovals) are distinguished from currency metabolites (pink ovals) in each reaction. If a reaction is reversible, then the links connecting the reaction to its reactant and product metabolites have arrows in both directions. (b) Unipartite metabolite graph representation for the three reactions in the glycolytic pathway. Note that before constructing the directed unipartite graph from the bipartite graph, we remove the currency metabolites.

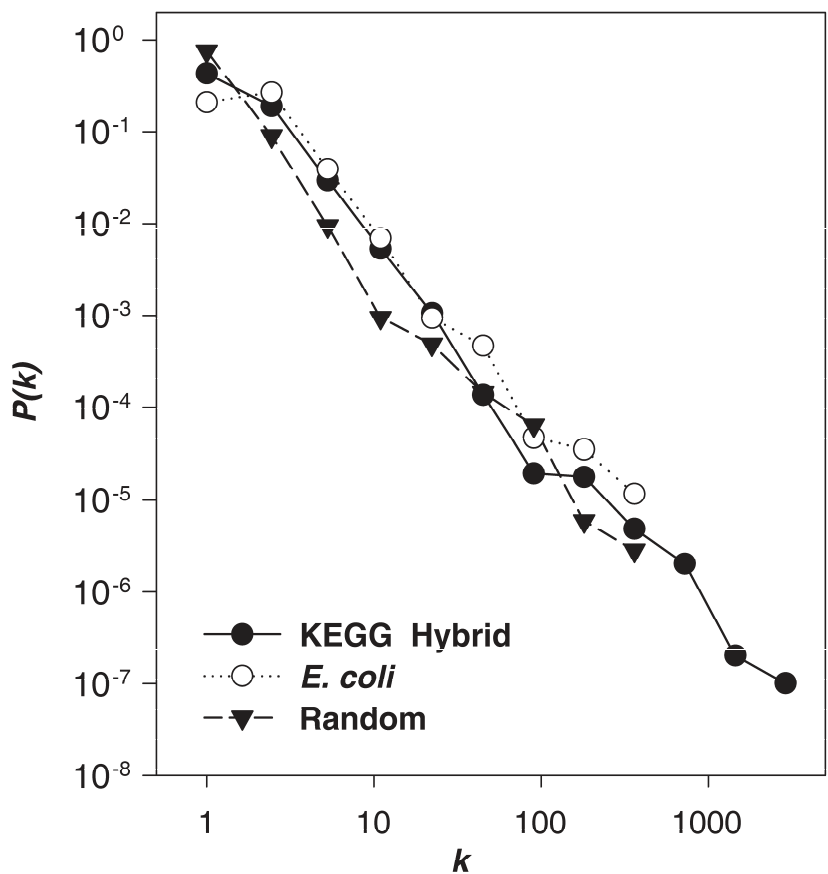

**Figure 2: The degree distribution of metabolites in metabolic networks.** The *x*-axis is the degree (*k*), the *y* axis is the probability *P(k)* that a metabolite has degree *k*. The relatively linear behavior on the log-log scale shows that a power law describes well the tail of the distribution. Three cases are shown. (1) All metabolites (and reactions) that are biochemically known (from the KEGG_Hybrid database), referred to as "KEGG_Hybrid". (2) Metabolites in the *E. coli* genome-scale metabolic model. (3) Metabolites in metabolic networks obtained by taking $n_E$ random reactions in KEGG_Hybrid, $n_E$ being the number of reactions in the *E. coli* genome-scale metabolic model (referred to as "Random"). In all cases, a power fit to the tail of the distribution leads to a satisfactory fit, the respective exponents being 2.31, 2.17 and 2.51.

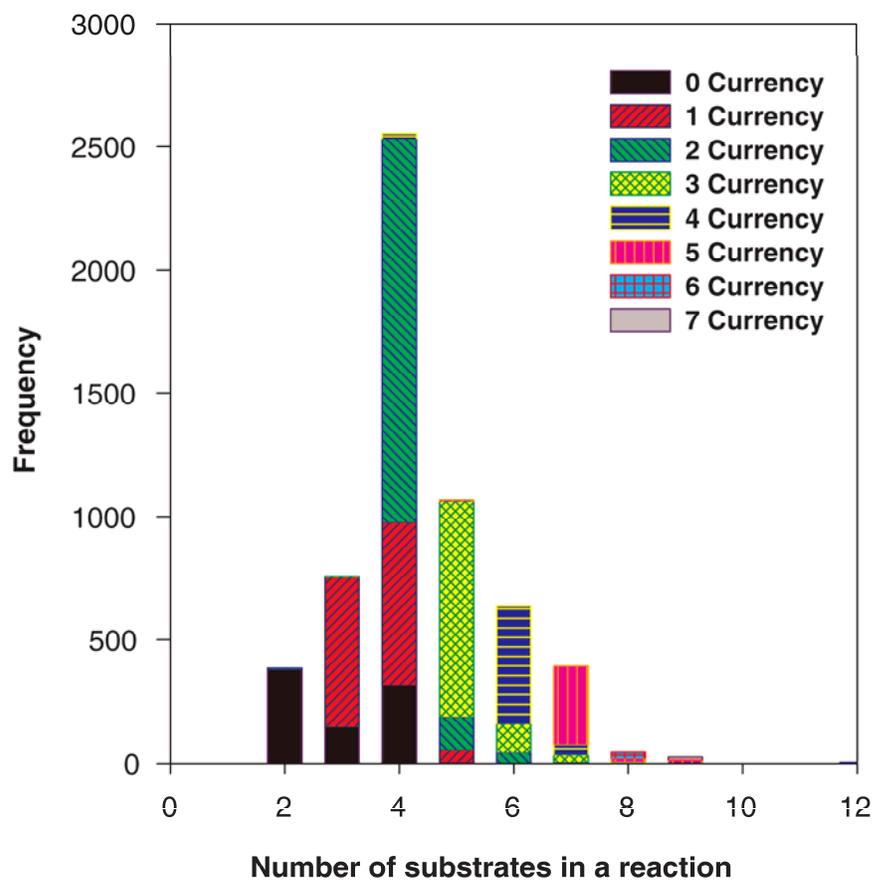

**Figure 3: The degree distribution for reactions in the KEGG_Hybrid universe of reactions.** For each degree (number of substrates in a reaction), we give the number of corresponding reactions having that degree. We also distinguish for each case the fraction of those reactions that have 0, 1, 2, … substrates which are currency metabolites.

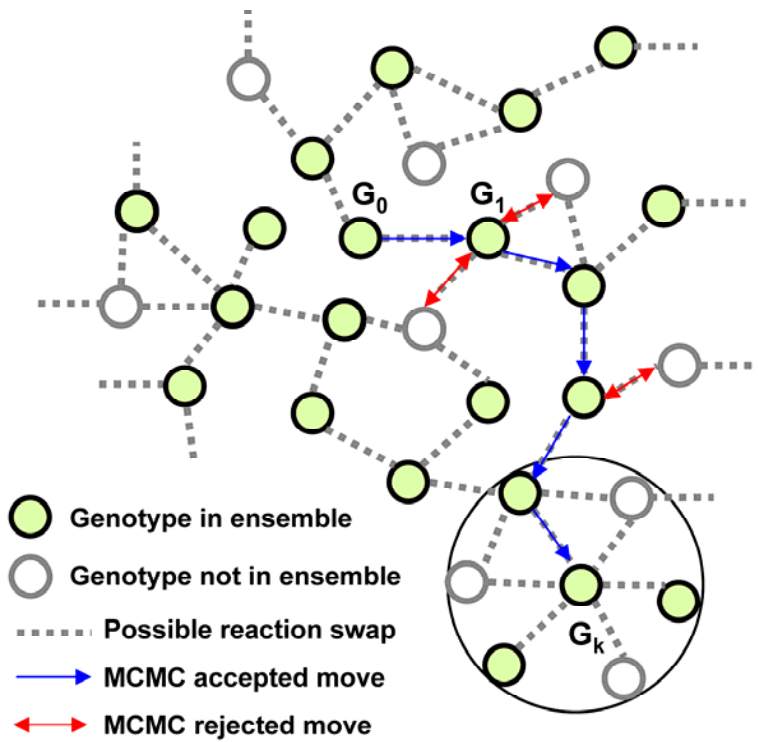

**Figure 4: MCMC sampling of genotypes in a randomized ensemble.** The space of genotypes with exactly $n_E$ reactions within KEGG_Hybrid (as in the *E. coli*) is very large, and only a tiny fraction of these genotypes are in any of the ensembles *RM, uRM, uRM-V1 etc*. MCMC allows one to sample this tiny fraction by generating a random walk restricted to the genotypes in the ensemble of interest. The MCMC starts with the *E. coli* genotype (shown in the figure as $G_0$) and proceeds as follows. At each trial step, a modified genotype is generated by applying a reaction swap to the current genotype. If the modified genotype satisfies the constraints of the ensemble, the trial move is accepted (shown in the figure as blue arrow) with the modified genotype becoming the next genotype of the walk. If the modified genotype does not satisfy the constraints of the ensemble, the trial step is rejected (shown in the figure by red arrows) and the walk stays at the previous genotype for that step. The advantage of using reaction swaps in our approach is that it leaves the number of reactions constant over time. The genotypes on the boundary of the large circle are in the neighborhood of genotype $G_k$ and differ from it by a single reaction swap.

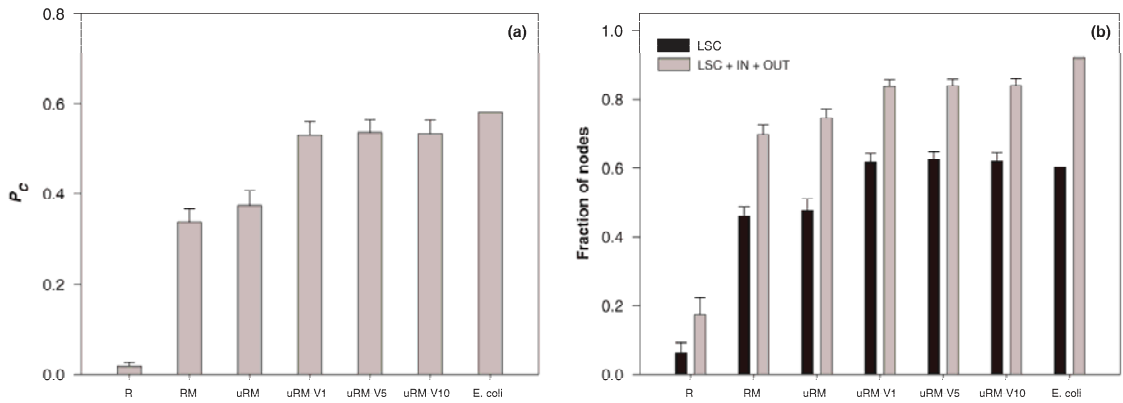

**Figure 5: Graph-based characteristics of the metabolic networks in the different ensembles.** (a) Probability $P_C$ that a path exists between two nodes taken at random in the directed metabolite-metabolite graph. (b) Fraction of nodes belonging to the largest strong component (LSC) or to the union of LSC, IN and OUT components. Different bars from left to right correspond to network ensembles incorporating an increasing number of constraints and the last bar corresponds to the *E. coli* metabolic network. The standard deviation is also displayed for each ensemble.

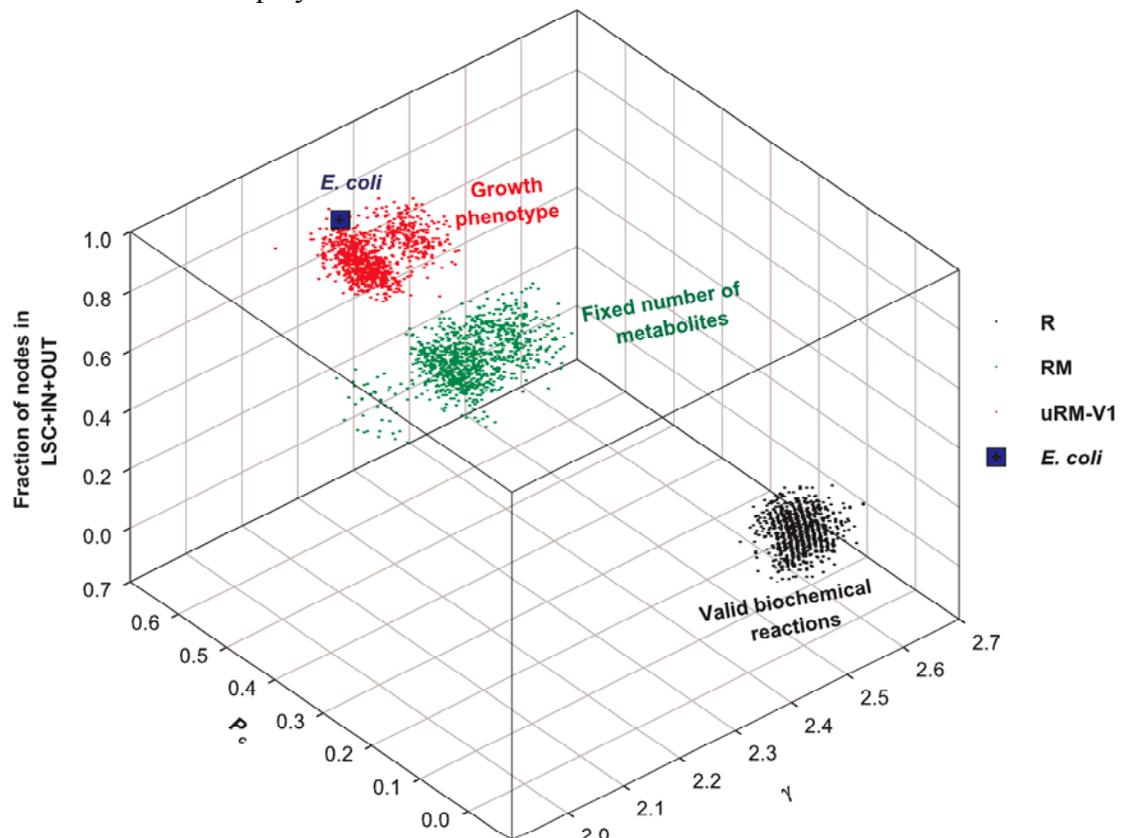

**Figure 6: Synthetic view of the statistical properties of randomized networks in different ensembles and comparison to the *E. coli* metabolic network.** The three axes are associated with graph characteristics of the networks. $P_C$ is the probability that a path exists between two nodes. $\gamma$ is the exponent of the power law fit to the metabolite degree distribution. The vertical axis is the fraction of nodes in the union of LSC, IN and OUT

components. Each cloud represents 1000 randomized networks in the ensemble considered.

## Supporting Information

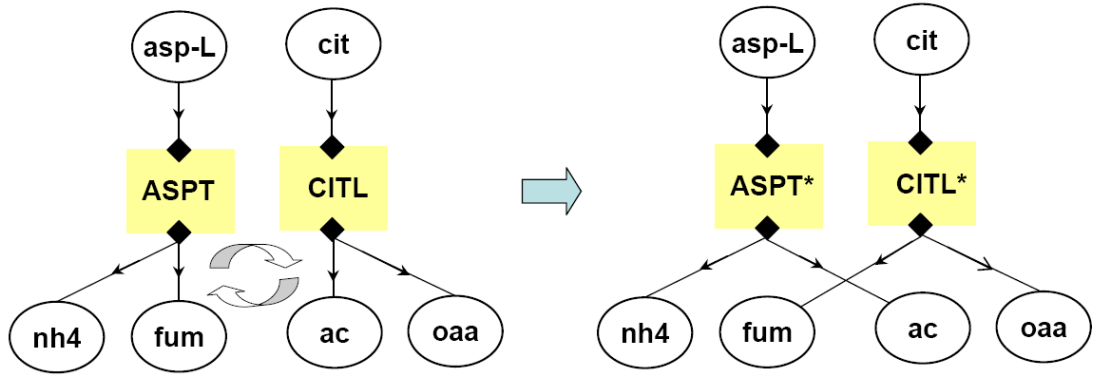

**Figure S1: Edge exchange randomization is biochemically meaningless.** The commonly used edge exchange or link permutation procedure for randomizing biological networks is inappropriate for metabolic networks as the method generates fictitious reactions violating balance of mass, charge and atomic elements. Here, starting with two reactions (ASPT: asp-L → fum + nh4; CITL: cit → oaa + ac), we perform an edge exchange associated to metabolites fum and ac that generates two new hypothetical reactions (ASPT*: asp-L → ac + nh4; CITL*: cit → oaa + fum) that violate balance of mass and atomic elements. Note that ac has 2 carbon atoms and fum has 4 carbon atoms.

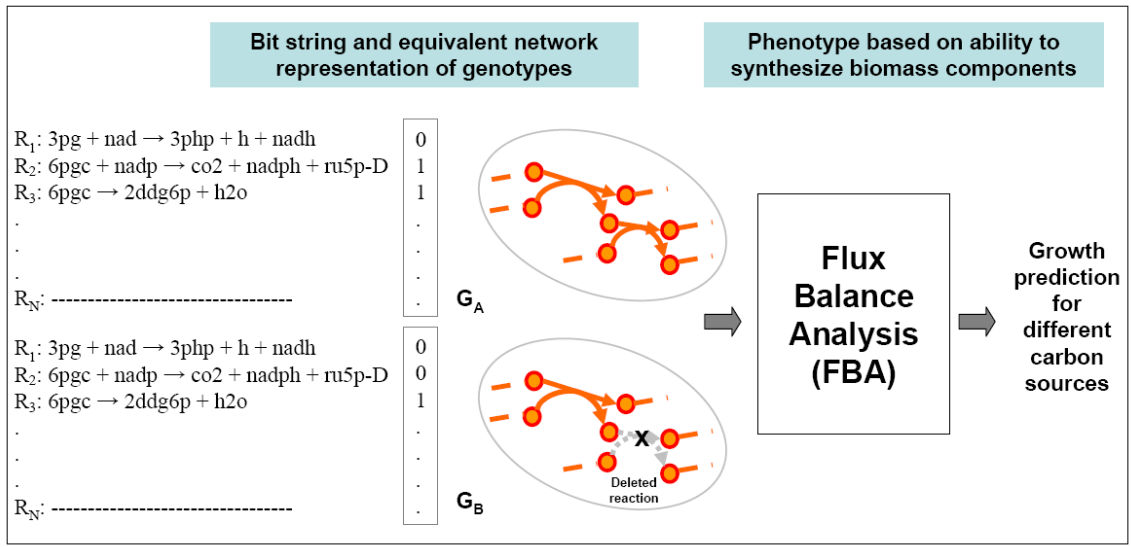

**Figure S2: Schematic summary of the relationship between genotypes and phenotypes.** The genotype specifies the list of reactions in a metabolic network. The phenotype is determined by whether the metabolic network can produce biomass components (growth) in a choice of chemical environments; this condition is computed using FBA.

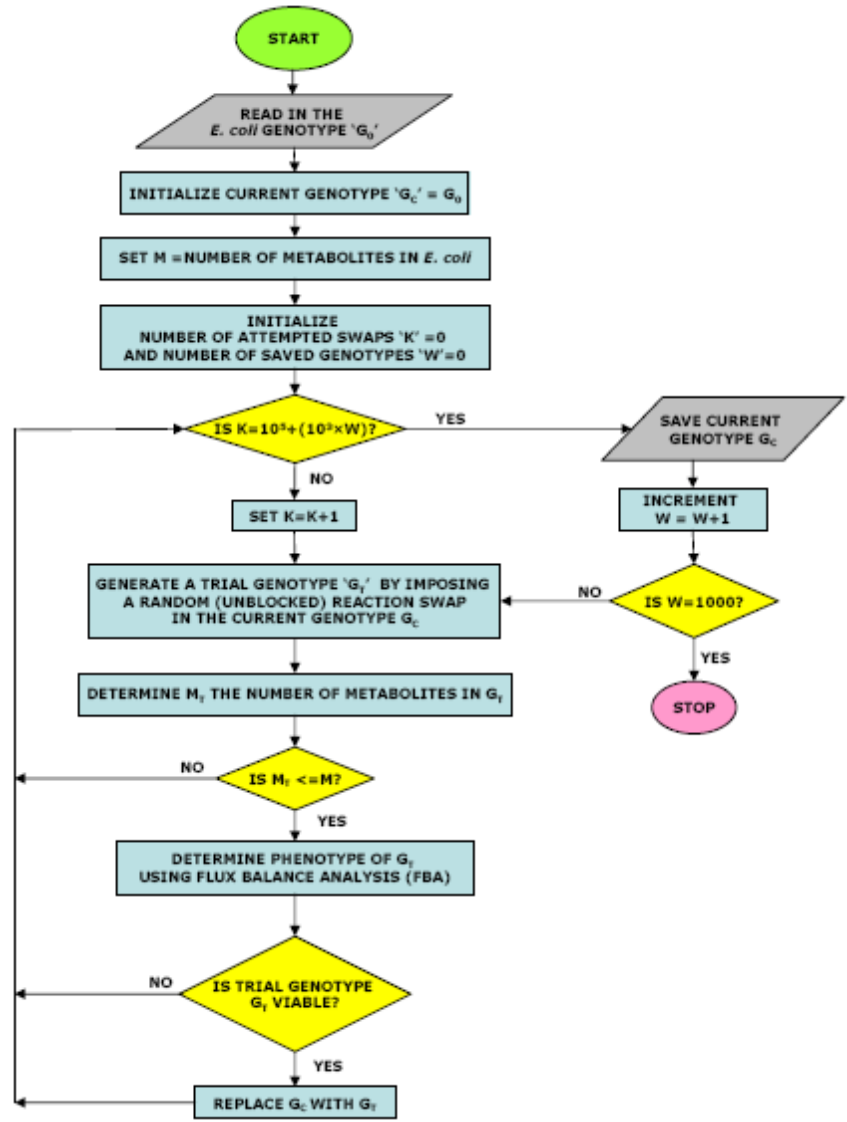

**Figure S3: Flowchart of the MCMC algorithm to sample genotypes in the ensemble uRM-V1.** The Markov chain starts with the *E. coli* genotype. We then perform $10^5$ Markov Chain steps to erase the memory of the initial genotype. After this initial phase, we continue the MCMC procedure to sample the genotype network and save every 1000$^{th}$ genotype generated. We terminate the Markov chain after saving 1000 genotypes. Note that the length of the run (and the choice of saving frequency) should be long enough to obtain a meaningful and uncorrelated sample of genotypes using this algorithm.

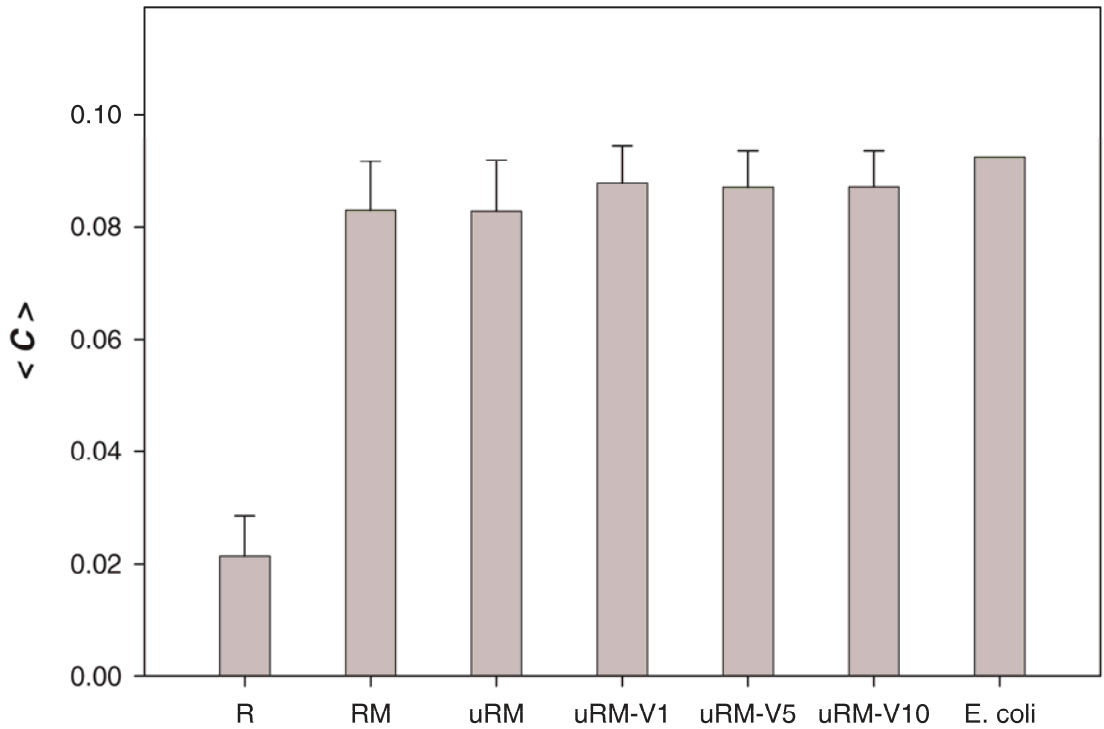

**Figure S4: Clustering coefficient *C* of the metabolic networks in the different ensembles.** Different bars from left to right correspond to network ensembles incorporating an increasing number of constraints and the last bar corresponds to the *E. coli* metabolic network. The standard deviation is also displayed for each ensemble.

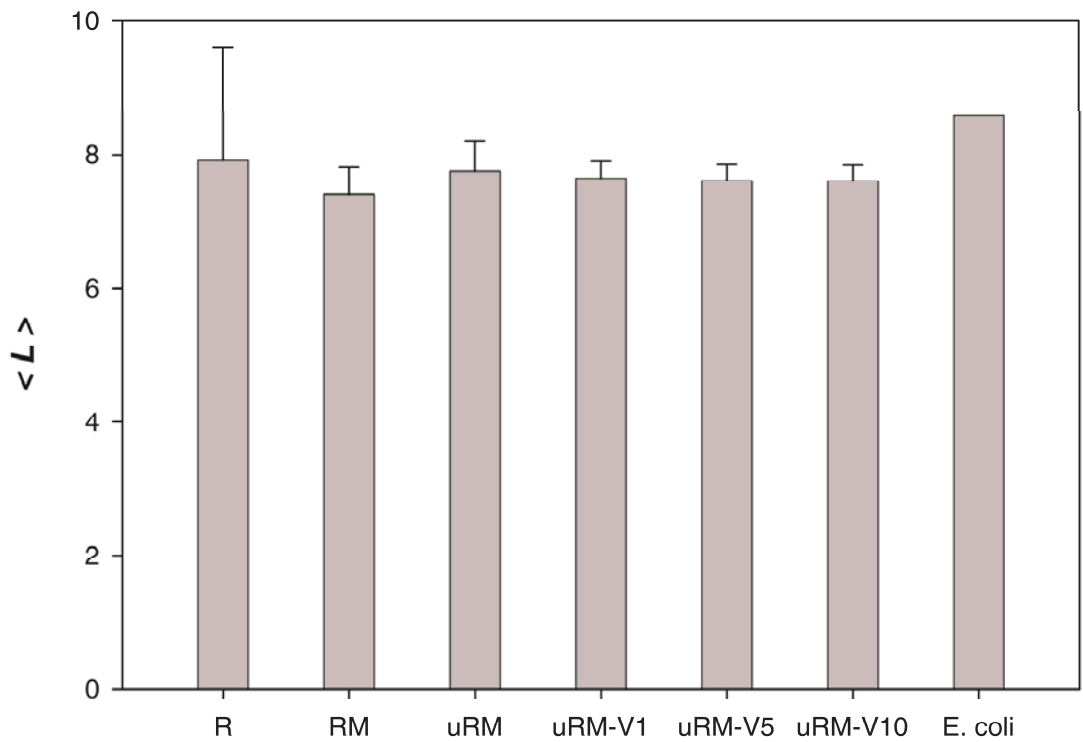

**Figure S5: Average path length *<L>* of the metabolic networks in the different ensembles.** Different bars from left to right correspond to network ensembles incorporating an increasing number of constraints and the last bar corresponds to the *E. coli* metabolic network. The standard deviation is also displayed for each ensemble.

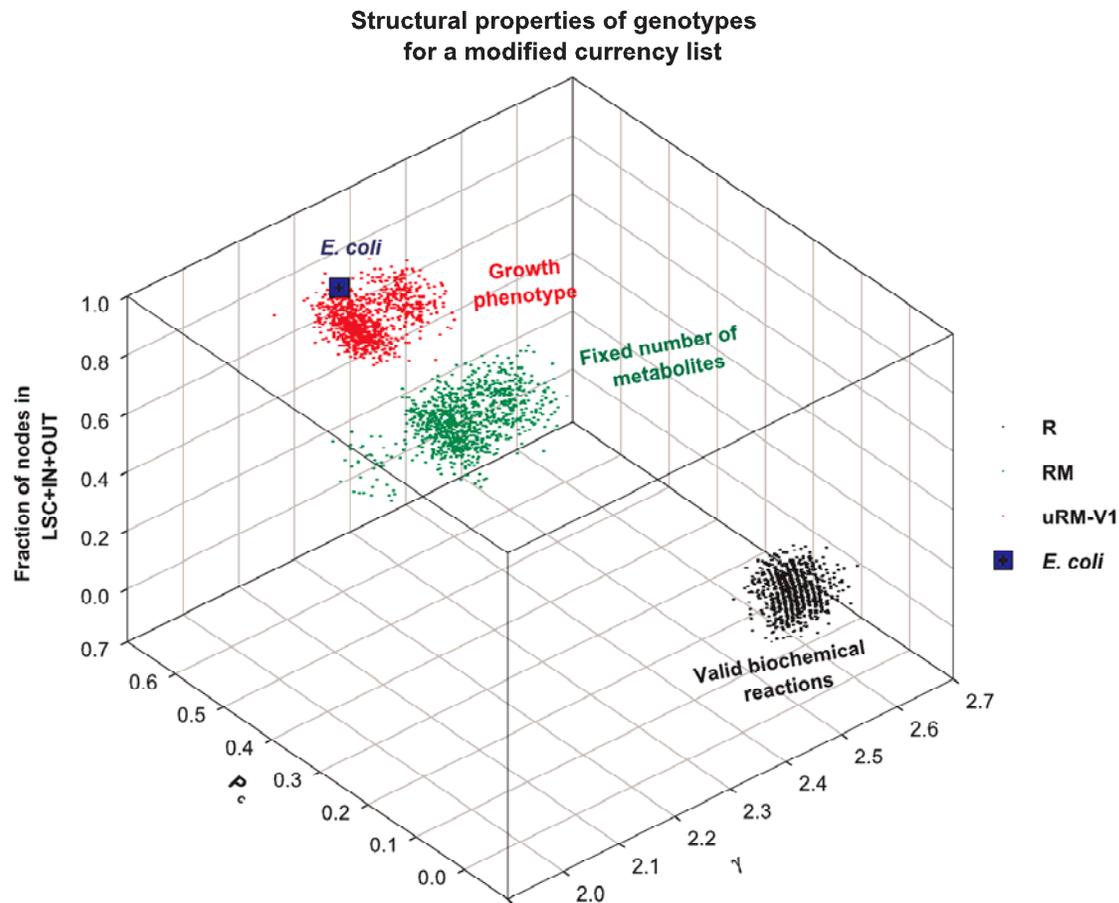

**Figure S6: Statistical properties of randomized networks in different ensembles and the *E. coli* metabolic network using a modified currency list.** The three axes are associated with graph characteristics of the networks and are same as in Figure 6. Each cloud represents 1000 randomized networks in the ensemble considered. In order to compute graph characteristics of randomized networks shown in this figure, we have constructed the metabolite-metabolite graph corresponding to each randomized network using a currency list modified from that listed in Table S1. The modified currency list was generated as follows. We first ranked metabolites in the currency list (given in Table S1) based on metabolite degree in the complete reaction database. The lowest degree metabolite in the currency list was designated rank 1. The 20 metabolites of smallest rank in this ranked currency list were then eliminated to generate the modified currency list used for computing graph characteristics shown in this figure. By comparing this figure with its analog (Figure 6), one sees that our conclusions are the same for the two definitions of currency metabolites.

.

**Table S1: List of currency metabolites used to construct the unipartite graph.** This list was built mostly from information in the paper by Ma and Zeng (Bioinformatics, 19, 270 (2003)). The metabolites shaded in grey were absent in the modified currency list used to generate Figure S6.

| Metabolite(s) | Context in which metabolite is regarded as currency |
|---|---|
| $H_2O$ | Always |
| $H_2O[e]$ | Always |
| $H^+$ | Always |
| $H^+[e]$ | Always |
| $O_2$ | Always |
| $O_2[e]$ | Always |
| $H_2O_2$ | Always |
| $CO_2$ | Always |
| $CO_2[e]$ | Always |
| $H_2CO_3$ | Always |
| $NH_3$ | Always |
| Nitrite | Always |
| Nitrite[e] | Always |
| Nitrate | Always |
| Nitrate[e] | Always |
| Nitric oxide | Always |
| $H_2S$ | Always |
| Sulfate | Always |
| Sulfite | Always |
| Sodium | Always |
| Sodium[e] | Always |
| Potassium | Always |
| Potassium[e] | Always |
| $Fe^{2+}$ | Always |
| $Fe^{2+}[e]$ | Always |
| $Fe^{3+}$ | Always |
| Orthophosphate | Always |
| Pyrophosphate | Always |
| CoA | Always except reactions in CoA synthesis pathway |
| Acetate | Always |
| ATP/ADP | Phosphate transfer |
| ATP/AMP | Phosphate transfer |
| ADP/AMP | Phosphate transfer |
| GTP/GDP | Phosphate transfer |
| GTP/GMP | Phosphate transfer |
| GDP/GMP | Phosphate transfer |
| CTP/CDP | Phosphate transfer |
| CTP/CMP | Phosphate transfer |
| CDP/CMP | Phosphate transfer |
| UTP/UDP | Phosphate transfer |
| UTP/UMP | Phosphate transfer |
| UDP/UMP | Phosphate transfer |
| ITP/IDP | Phosphate transfer |
| ITP/IMP | Phosphate transfer |
| IDP/IMP | Phosphate transfer |
| NADH/NAD$^+$ | Hydrogen transfer |
| NADPH/NADP$^+$ | Hydrogen transfer |
| FADH/FAD$^+$ | Hydrogen transfer |
| Reduced Acceptor/Acceptor | Hydrogen transfer |
| Glutathione/Oxidized Glutathione | Hydrogen transfer |
| Ferrocytochrome c/Ferricytochrome c | Hydrogen transfer |
| Reduced ferredoxin/Oxidized ferredoxin | Hydrogen transfer |
| Reduced rubredoxin/Oxidized rubredoxin | Hydrogen transfer |
| Reduced adrenal ferredoxin/Oxidized adrenal ferredoxin | Hydrogen transfer |
| Dihydrobiopterin/Tetrahydrobiopterin | Hydrogen transfer |
| Ubiquinol-8/Ubiquinone-8 | Hydrogen transfer |
| 5,10-Methylenetetrahydrofolate/Tetrahydrofolate | One carbon unit transfer |
| 10-Formyltetrahydrofolate/Tetrahydrofolate | One carbon unit transfer |
| 5-Methyltetrahydrofolate/Tetrahydrofolate | One carbon unit transfer |
| 5,10-Methenyltetrahydrofolate/Tetrahydrofolate | One carbon unit transfer |
| 5-Formiminotetrahydrofolate/Tetrahydrofolate | One carbon unit transfer |
| 10-Formyldihydrofolate/Tetrahydrofolate | One carbon unit transfer |
| 5-Formyltetrahydrofolate/Tetrahydrofolate | One carbon unit transfer |
| S-Adenosyl-L-methionine/S-Adenosyl-L-homocysteine Adenosine | Methyl group transfer |
| 3',5'-bisphosphate/3'-Phosphoadenylyl sulfate | Sulfate group transfer |
| UDP-glucose/UDP | Monosaccharide unit transfer |
| UDP-N-acetyl-D-glucosamine/UDP | Monosaccharide unit transfer |
| UDP-D-galactose/UDP | Monosaccharide unit transfer |
| UDP-glucuronate/UDP | Monosaccharide unit transfer |
| UDP-D-xylose/UDP | Monosaccharide unit transfer |
| UDP-N-acetyl-D-galactosamine/UDP | Monosaccharide unit transfer |
| Glutamate/Oxoglutarate | Amino group transfer |
| Glutamine/Glutamate | Amino group transfer |
| Pyruvate/Alanine | Amino group transfer |
| Pyruvate/Phosphoenolpyruvate | Phosphate transfer |
| Acetyl-CoA/CoA | Acetyl group transfer |